\documentclass[conference]{IEEEtran}

\usepackage{amsmath,amssymb,amsfonts,amsthm}
\usepackage{mathtools}
\usepackage{bm}

\usepackage{graphicx}
\usepackage{booktabs}
\usepackage[table]{xcolor}
\usepackage{colortbl}
\usepackage{tikz}
\usetikzlibrary{positioning,calc}

\usepackage{algorithm}
\usepackage{algpseudocode}
\algrenewcommand\algorithmicrequire{\textbf{Input:}}
\algrenewcommand\algorithmicensure{\textbf{Output:}}

\usepackage{dsfont}
\usepackage{url}
\usepackage{microtype}

\usepackage{hyperref}
\hypersetup{colorlinks,allcolors=black}

\begin{document}


\title{Conflict-Free Color-Clustered Sequential Belief-Propagation Decoding of Quantum LDPC Codes via Reinforcement Learning\vspace{-0.15cm}}


\author{
Mohsen~Moradi$^*$,
Taejoon~Kim$^*$,
and
R\'emi~A.~Chou$^\dag$\\
$^*$School of Electrical, Computer and Energy Engineering,
Arizona State University, Tempe, AZ 85287, USA\\ 
Email: \{mmorad11, taejoonkim\}@asu.edu, \\
$^\dag$Department of Computer Science and Engineering, 
The University of Texas at Arlington, Arlington, TX, USA \\
Email: remi.chou@uta.edu
\vspace{-0.55cm}}

\maketitle

\begin{abstract}
Belief-propagation (BP) decoding for quantum low-density parity-check (QLDPC) codes is attractive due to its low complexity and low latency, but it is often limited by short cycles, degeneracy, and convergence failures. Reinforcement-learning-based sequential BP decoding (RL-S) improves BP by learning a syndrome-dependent variable-node (VN) update order, but its VN-by-VN schedule has limited within-iteration parallelism. In this paper, we propose a conflict-free color-clustered extension of RL-S. We construct a VN conflict graph in which two VNs are adjacent if they share an X-type or Z-type check, and color this graph so that same-color VNs have disjoint check neighborhoods. This also prevents VNs from the same Tanner 4- or 6-cycle from being updated simultaneously. During decoding, the trained VN-level Q-table selects a seed VN, and all remaining VNs with the same color are updated in parallel using the same pre-batch messages. For the \([[288,12,18]]\) bivariate-bicycle code over the depolarizing channel, our proposed decoder achieves block-error-rate performance close to VN-level RL-S while reducing the scheduling decisions from 288 VNs to 11 color classes per BP iteration.
\vspace{-.3cm}
\end{abstract}


\section{Introduction}

Quantum low-density parity-check (QLDPC) codes are a promising route toward
low-overhead quantum error correction because they combine sparse stabilizer
checks with favorable rate--distance scaling and strong finite-length
constructions
\cite{mackay2004sparse,breuckmann2021quantum,
panteleev2022asymptoticallygood,bravyi2024high}.  Their sparsity makes
belief propagation (BP) a natural low-complexity decoder.  However, standard
flooded BP is often unreliable on QLDPC Tanner graphs: stabilizer
commutativity creates many short cycles, while quantum degeneracy allows
many Pauli errors to share the same syndrome or logical action.  These
effects can cause oscillation, non-convergence, or convergence to a
syndrome-consistent but logically incorrect estimate
\cite{poulin2008iterative,kuo2022exploiting}.

A direct way to improve BP without adding a heavy post-processing stage,
such as BP with ordered-statistics decoding (BP-OSD)
\cite{panteleev2021degenerate,roffe2020decoding}, is to change the
message-update schedule. In classical LDPC decoding, serial and
layered schedules are known to improve convergence by reusing newly computed
messages within an iteration
\cite{hocevar2004layered,sharon2007serial}. 
For QLDPC codes, layered and random-order schedules, informed dynamic
scheduling, and fixed sequential check-node/variable-node schedules have shown that the update order can also substantially affect BP decoding
\cite{ducrest2023layered,huang2026informed,moradi2026sequential}.
Building on this idea, reinforcement-learning-based sequential BP decoding
(RL-S) learns a syndrome-dependent variable-node (VN) update order from local
residual-mismatch states while keeping the BP update equations unchanged
\cite{moradi2026rlqldpc}. 
A related list variant further improves reliability by exploring competing local Pauli decisions \cite{moradi2026rlls}.
The main drawback of VN-level RL-S is that one BP pass may require up to \(n\) sequential VN decisions, where \(n\) is the code block length, limiting within-iteration parallelism.

This paper proposes a conflict-free color-clustered extension of RL-S for
QLDPC decoding.  We form a VN conflict graph in which two VNs are adjacent
whenever they share at least one \(X\)-type or \(Z\)-type check, and then
color this graph so that same-color VNs have disjoint check neighborhoods.
During decoding, the trained VN-level RL policy selects a seed VN, and all
remaining VNs with the same color as the seed are updated in parallel using
the same pre-batch messages.  
The construction is related in spirit to conflict-aware layered and clustered schedules for classical and fair-density parity-check codes, including graph-coloring-based layered scheduling and learned clustered check-node scheduling
\cite{dupraz2018lowlatency,habib2021clustered,hosseinzadeh2025layered},
but here the coloring is applied to the joint \(H_X/H_Z\) VN conflict graph of a QLDPC code and is used to parallelize a learned VN-level QLDPC BP schedule.
The clustered RL-S decoder of \cite{moradi2026RL_Cluster} learns a cluster-level schedule over a fixed random VN partition; in contrast, our method reuses the trained VN-level RL-S policy and exploits the joint Tanner-graph structure induced by \(H_X\) and \(H_Z\) to form conflict-free color classes with disjoint check neighborhoods.

\section{Preliminaries and Decoder Model}
\label{sec:background}

We consider a Calderbank–Shor–Steane (CSS) QLDPC code \cite{calderbank1996good,steane1996error} specified by sparse binary matrices
\(H_X\in\mathbb F_2^{m_X\times n}\) and
\(H_Z\in\mathbb F_2^{m_Z\times n}\), with
\(H_XH_Z^{\mathsf T}=0\).  For a Pauli error
\(q=(q_1,\ldots,q_n)\), define
\[
e^X_i=\mathds{1}\{q_i\in\{X,Y\}\},\qquad
e^Z_i=\mathds{1}\{q_i\in\{Z,Y\}\}.
\]
The measured syndrome is
\(
s^Z=H_Z e^X,~ s^X=H_X e^Z,
\)
where all operations are over \(\mathbb F_2\).  Given a current Pauli estimate
\(\hat q\), let \(\hat e^X\) and \(\hat e^Z\) denote its binary components and
define the residual mismatch vectors
\(
\delta^Z=s^Z\oplus H_Z\hat e^X,~
\delta^X=s^X\oplus H_X\hat e^Z .
\)
A check is satisfied when the corresponding residual bit is zero.  Under
depolarizing noise, the \(X\)- and \(Z\)-components are coupled through \(Y\)
errors; we therefore use the same quaternary/two-stream BP update model and
the same trained VN-level RL-S Q-table as in
\cite{moradi2026rlqldpc}.  In RL-S, each candidate VN $v_i$ is
assigned a local state \(\sigma_i\) from the residual patterns around that VN,
and the next VN is selected according to
\[
v^\star \in \arg\max_{v_i\in\mathcal R} Q(\sigma_i,i),
\]
where \(\mathcal R\) is the set of VNs not yet visited in the current BP pass.
Our proposed decoder keeps this learned seed selection rule but replaces the
single selected VN by a conflict-free same-color batch.

\section{VN-Color-Guided Parallelization of RL-S}
\label{subsec:vn_color_rl_qsvns}

The original RL-S decoder updates one variable node (VN) at a time, which
creates a serial scheduling depth of order \(n\) per BP iteration. To increase within-iteration parallelism while
reusing the trained VN-level RL policy, we introduce a coloring-based grouping
of VNs.

The key idea is to update several VNs in parallel only when their updates
do not directly conflict through a common check node. For this purpose, we
define the VN conflict graph
\[
G_{\mathrm{VN}}=(\mathcal V,\mathcal E_{\mathrm{VN}}),
\qquad
\mathcal V=\{v_1,\ldots,v_n\},
\]
where each vertex \(v_i\) represents one physical qubit/VN. Two distinct
VNs \(v_i\) and \(v_{i'}\), with \(i\neq i'\), are adjacent in
\(G_{\mathrm{VN}}\) if they share at least one check node in either Tanner
graph:
\[
\begin{aligned}
(v_i,v_{i'})\in \mathcal E_{\mathrm{VN}}&
\Longleftrightarrow\;
\exists c_j \text{ such that } H_X(j,i)=H_X(j,i')=1 \\
&~~~~\text{or } 
\exists c_j \text{ such that } H_Z(j,i)=H_Z(j,i')=1 .
\end{aligned}
\]
Equivalently, if
\(
\widetilde A_{\mathrm{VN}}
=
\mathds{1}\!\left\{
H_X^{\mathsf T}H_X+H_Z^{\mathsf T}H_Z>0
\right\},
\)
then the adjacency matrix of the VN conflict graph is obtained by removing
self-conflicts:
\(
A_{\mathrm{VN}}
=
\widetilde A_{\mathrm{VN}}
-
\operatorname{diag}(\widetilde A_{\mathrm{VN}}).
\)

We then compute a proper coloring
\[
\kappa:\mathcal V\rightarrow \{1,\ldots,\chi_{\mathrm{VN}}\}
\]
such that
\[
(v_i,v_{i'})\in \mathcal E_{\mathrm{VN}}
\quad \Longrightarrow \quad
\kappa(v_i)\neq \kappa(v_{i'}).
\]
Hence, if two VNs have the same color, then they do not share any
\(X\)-type or \(Z\)-type check node. Equivalently,
\[
\kappa(v_i)=\kappa(v_{i'})
\quad \Longrightarrow \quad
\mathcal N_X(v_i)\cap \mathcal N_X(v_{i'})=\emptyset
\]
and
\(
\mathcal N_Z(v_i)\cap \mathcal N_Z(v_{i'})=\emptyset,
\)
where
\[
\begin{aligned}
\mathcal N_X(v_i)
&\triangleq \{c_j:H_X(j,i)=1\},\\
\mathcal N_Z(v_i)
&\triangleq \{c_j:H_Z(j,i)=1\}.
\end{aligned}
\]
Therefore, VNs with the same color have disjoint check-node neighborhoods.
This guarantees that all VN updates within a single color class can be
computed in parallel without direct check-node collisions.

The same coloring rule also prevents a single color batch from containing
multiple VNs that form a Tanner 4-cycle or a Tanner 6-cycle. To see this,
consider first a Tanner 4-cycle
\[
v_i - c_j - v_{i'} - c_{j'} - v_i .
\]
In this cycle, the two VNs \(v_i\) and \(v_{i'}\) share the check node
\(c_j\), and they also share the check node \(c_{j'}\). Hence,
\((v_i,v_{i'})\in\mathcal E_{\mathrm{VN}}\). Since the coloring is proper,
they cannot receive the same color:
\(
\kappa(v_i)\neq \kappa(v_{i'}).
\)
Thus, one color batch can contain at most one of the two VNs involved in
such a 4-cycle.

Similarly, consider a Tanner 6-cycle
\[
v_{i_1}-c_{j_1}-v_{i_2}-c_{j_2}-v_{i_3}-c_{j_3}-v_{i_1}.
\]
Here, \(v_{i_1}\) and \(v_{i_2}\) share \(c_{j_1}\), \(v_{i_2}\) and
\(v_{i_3}\) share \(c_{j_2}\), and \(v_{i_3}\) and \(v_{i_1}\) share
\(c_{j_3}\). Therefore, the three VNs form a triangle in the VN conflict
graph:
\[
(v_{i_1},v_{i_2}),\quad
(v_{i_2},v_{i_3}),\quad
(v_{i_3},v_{i_1})
\in \mathcal E_{\mathrm{VN}}.
\]
A proper coloring must assign different colors to all three VNs. Hence,
one color batch can contain at most one VN from this 6-cycle. 
This means that the proposed color-class update does not update
multiple VNs from the same 4-cycle or 6-cycle simultaneously.

For longer cycles, this no-common-check condition is not sufficient to
separate all VNs that belong to the same cycle. For example, consider the
Tanner 8-cycle
\[
v_{i_1}-c_{j_1}-v_{i_2}-c_{j_2}
-v_{i_3}-c_{j_3}-v_{i_4}-c_{j_4}-v_{i_1}.
\]
The adjacent VN pairs in this cycle share check nodes, so the VN conflict
graph contains the edges
\(
(v_{i_1},v_{i_2}),~
(v_{i_2},v_{i_3}),~
(v_{i_3},v_{i_4}),~
(v_{i_4},v_{i_1}).
\)
However, the non-adjacent pairs \(v_{i_1}\) and \(v_{i_3}\), and
\(v_{i_2}\) and \(v_{i_4}\), do not necessarily share any check node.
Therefore, these pairs are not necessarily adjacent in
\(G_{\mathrm{VN}}\), and a proper coloring may assign them the same color.
Consequently, a single color batch may contain, for example,
\(v_{i_1}\) and \(v_{i_3}\), even though both VNs belong to the same
8-cycle.

In our proposed color-guided version, we first select the same seed VN
\(v^\star\) from the Q-table, and then use its color
\(
\ell^\star=\kappa(v^\star)
\)
to define the same-color candidate set
\[
S_{\ell^\star}
=
\{v_i\in \mathcal R:\kappa(v_i)=\ell^\star\}.
\]
The selected parallel batch is the entire remaining same-color set:
\(
\mathcal B_t = S_{\ell^\star}.
\)

All VNs in \(\mathcal B_t\) are then updated in parallel using the same
pre-batch messages, and the batch is removed from the remaining set:
\(
\mathcal R \leftarrow \mathcal R\setminus \mathcal B_t.
\)
Since all VNs in \(\mathcal B_t\) have the same color, they have disjoint
check-node neighborhoods. Therefore, their message updates can be computed
simultaneously without direct check-node conflicts.

This procedure lifts the already trained VN-level RL policy to a
color-class parallel schedule. The RL policy is still used to select the
seed VN, which determines the next color class to process, but once the
color is selected, all remaining VNs of that color are updated together.
Consequently, each BP iteration requires at most
\(\chi_{\mathrm{VN}}\) batch-scheduling decisions, rather than \(n\)
individual VN-scheduling decisions.

Our procedure is summarized in
Algorithm~\ref{alg:color_guided_rl_qsvns}.  The VN conflict graph and its
coloring can be computed offline.  During online decoding, the trained VN-level Q-table is used only to
select the seed VN; the selected seed then determines the conflict-free color
class that is updated in parallel.

\begin{algorithm}[t]
\caption{Color-Class-Guided RL-S Inference}
\label{alg:color_guided_rl_qsvns}
\begin{algorithmic}[1]
\Require Parity-check matrices \(H_X,H_Z\), trained VN-level Q-table \(Q\),
maximum number of iterations \(T\)
\Ensure Estimated Pauli error \(\hat q\), or a non-convergence flag

\State Compute
\(
\widetilde A_{\mathrm{VN}}
=
\mathds{1}\!\left\{
H_X^{\mathsf T}H_X+H_Z^{\mathsf T}H_Z>0
\right\}.
\)
\State Set
\(
A_{\mathrm{VN}}
=
\widetilde A_{\mathrm{VN}}
-
\operatorname{diag}(\widetilde A_{\mathrm{VN}}).
\)

\State Compute a proper coloring \(\kappa(v_i)\) of \(G_{\mathrm{VN}}\).
\State Initialize BP messages, hard decisions, residual mismatches, and
local VN states.

\For{\(t=1,\ldots,T\)}
    \State \(\mathcal R\leftarrow \{v_1,\ldots,v_n\}\)
    \While{\(\mathcal R\neq \emptyset\)}
        \If{the residual syndrome is zero}
            \State \Return \(\hat q\)
        \EndIf

        \State Compute RL scores
        \[
        g_{v_i}=Q(\sigma_{v_i},v_i), \qquad v_i\in\mathcal R.
        \]
        \State Select the seed VN:
        \(
        v^\star=\arg\max_{v_i\in\mathcal R} g_{v_i}.
        \)
        \State Let \(\ell^\star=\kappa(v^\star)\).
        \State Form the same-color candidate set
        \[
        S_{\ell^\star}
        =
        \{v_i\in\mathcal R:\kappa(v_i)=\ell^\star\}.
        \]
        \State Select the full color-class batch
        \(
        \mathcal B_t = S_{\ell^\star}.
        \)
        \State Update all VNs in \(\mathcal B_t\) in parallel using the
        pre-batch messages.
        \State Commit the message updates, hard-decision changes, residual
        mismatch updates, and local-state updates.
        \State \(\mathcal R\leftarrow \mathcal R\setminus \mathcal B_t\).
    \EndWhile
\EndFor

\State \Return non-convergence with final estimate \(\hat q\)
\end{algorithmic}
\end{algorithm}

\section{Numerical Results}
The block error rate (BLER) is used as the main performance metric. In each
Monte-Carlo trial, a decoding failure is declared if the decoder either does not
converge to the measured syndrome within the prescribed iteration cap, or if the
resulting syndrome-consistent correction differs from the channel error by a
nontrivial logical operator. 
Throughout this section, \(p\) denotes the depolarizing error probability.

\begin{figure}[t]
  \centering
  \includegraphics[width=\linewidth]{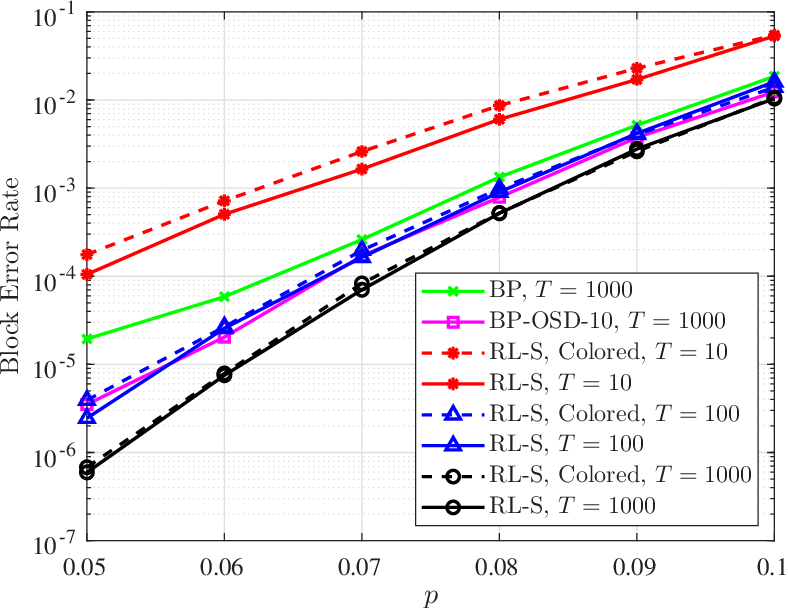}
  \caption{Block-error-rate performance comparison between our proposed color-clustered RL-S decoder and the VN-level RL-S decoder for the \( [[288,12,18]] \) BB code over the depolarizing channel.\vspace{-.5cm}}
  \label{fig:BB288_RL_LS}
\end{figure}

Fig.~\ref{fig:BB288_RL_LS} shows the BLER performance of the
\([[288,12,18]]\) bivariate-bicycle (BB) code~\cite{bravyi2024high} over the
depolarizing channel. We compare the VN-level RL-S decoder with the proposed
color-clustered RL-S decoder. 
As references, we also include standard BP and BP-OSD-10
\cite{panteleev2021degenerate,roffe2020decoding}, where the BP stage is run with a maximum of \(T=1000\) iterations and the OSD order is 10.

The results show that color clustering preserves most of the performance gain
of the fully sequential VN-level RL-S schedule. In particular, for small iteration caps, updating an entire color class at once introduces a modest performance loss relative to VN-by-VN RL-S, as expected, because the schedule provides less control over the individual VN update order. However, this gap decreases as the iteration cap increases.
At \(T=1000\), the color-clustered RL-S curve remains close to the VN-level RL-S curve over the simulated range. Moreover, already at \(T=100\), the proposed decoder achieves BLER performance comparable to
BP-OSD-10, while avoiding the additional OSD post-processing stage.

For this code, the VN conflict graph is colored with
\(\chi_{\mathrm{VN}}=11\) colors. Hence, one color-clustered BP pass requires
at most 11 sequential color-layer updates, instead of 288 sequential VN updates
in the original RL-S schedule. Therefore, with \(T=100\), the worst-case
sequential scheduling depth of the proposed decoder is bounded by
$1100$
color-layer updates, compared with \(288\times 100=28800\) VN-level scheduling
steps for VN-level RL-S at the same iteration cap. This corresponds to about a
\(288/11 \approx 26.2\)-fold reduction in sequential scheduling depth per BP
pass. The actual number of color-layer updates can be smaller because early stopping is applied once the residual syndrome vanishes.

\begin{figure}[t]
  \centering
  \includegraphics[width=\linewidth]{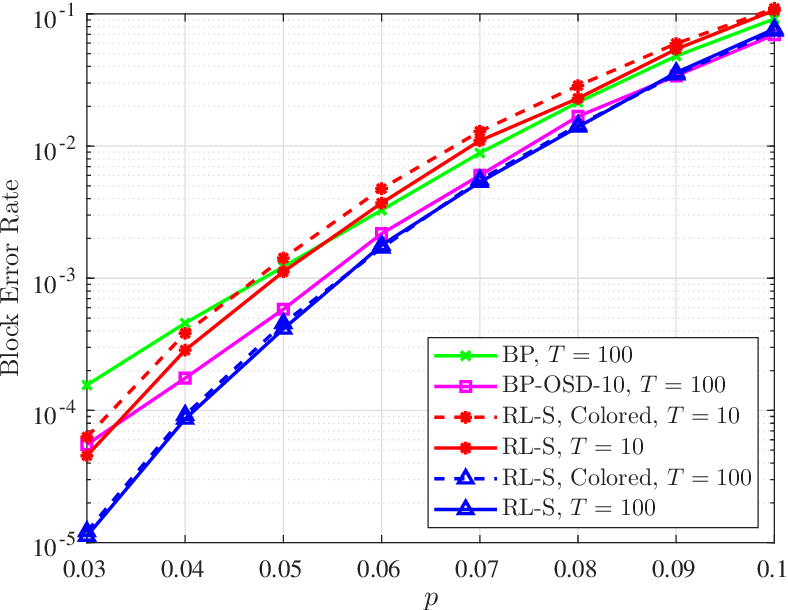}
  \caption{Block-error-rate performance comparison between our proposed color-clustered RL-S decoder and the VN-level RL-S decoder for the \( [[144,12,12]] \) BB code over the depolarizing channel.\vspace{-.3cm}}
  \label{fig:RLSvColor_BB144_FER}
\end{figure}

Fig.~\ref{fig:RLSvColor_BB144_FER} reports the corresponding results for the
\([[144,12,12]]\) BB code ~\cite{bravyi2024high}. The VN conflict graph for this code is colored with
\(\chi_{\mathrm{VN}}=9\) colors, so one color-clustered BP pass requires at
most 9 sequential color-layer updates instead of 144 VN updates. The proposed
decoder closely tracks the VN-level RL-S decoder for both tested iteration
caps, with only a small loss due to color-class batching. At \(T=100\), it is
also competitive with BP-OSD-10, while requiring at most \(9T=900\)
color-layer updates.

\begin{figure}[t]
  \centering
  \includegraphics[width=\linewidth]{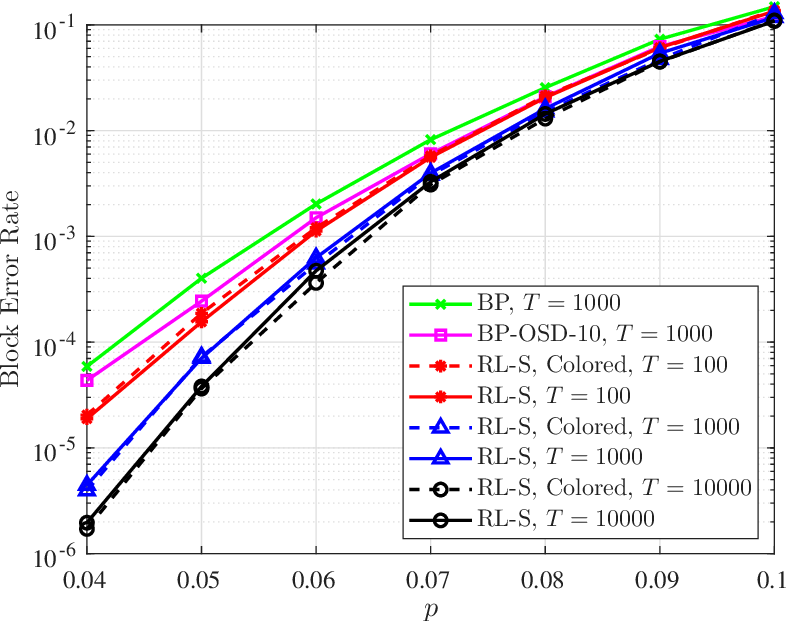}
  \caption{Block-error-rate performance comparison between our proposed
color-clustered RL-S decoder and the VN-level RL-S decoder for the
\([[180,10,15\leq d\leq18]]\) A5 code over the depolarizing channel.\vspace{-.3cm}}
  \label{fig:RLSvColor_A5_n180_FER}
\end{figure}

Fig.~\ref{fig:RLSvColor_A5_n180_FER} shows the results for the
\([[180,10,15\leq d\leq18]]\) A5 code~\cite{panteleev2021degenerate}. In this case,
\(\chi_{\mathrm{VN}}=17\), reducing the sequential scheduling depth from 180
VN-level decisions to at most 17 color-layer decisions per BP pass. As \(T\)
increases, the color-clustered RL-S curves remain close to the VN-level RL-S
curves and outperform standard BP over most of the simulated range. Thus, the
proposed batching preserves most of the RL-S gain while giving about
\(180/17\approx 10.6\)-fold fewer sequential scheduling decisions per pass.

\section{Conclusions}\label{sec:conclusion}

In this paper, we proposed a conflict-free color-clustered extension of
VN-level RL-S decoding for QLDPC codes. By coloring the VN conflict graph
induced jointly by \(H_X\) and \(H_Z\), same-color VNs have disjoint check
neighborhoods and can be updated in parallel from a common pre-batch message
state. The resulting decoder reuses the trained VN-level RL policy to select
color layers, reducing the scheduling depth from \(n\) VN decisions to
\(\chi_{\mathrm{VN}}\) color-layer decisions per BP pass. Numerical results on
several codes show that the proposed decoder preserves most of the BLER
performance of fully sequential RL-S while providing substantially more
within-iteration parallelism and avoiding OSD post-processing.

\section{Acknowledgment}
This work is in part supported by the National Science Foundation (NSF) under grants CNS2451268, CNS2514415, and ITE2515378, and the Office of Naval Research (ONR) under Grant N000142112472.

\bibliographystyle{IEEEtran}

\bibliography{bibliography}

\end{document}